\documentclass[conference]{IEEEtran}

\usepackage[utf8]{inputenc} 
\usepackage[T1]{fontenc}
\usepackage{url}
\usepackage{ifthen}
\usepackage{cite}
\usepackage[cmex10]{amsmath} %
\usepackage{mathtools}

\usepackage{amsfonts}
\usepackage{thmtools}

\usepackage[
	ruled,
	vlined,
	boxed, 
	linesnumbered, 
	commentsnumbered]{algorithm2e}

\usepackage{siunitx}

\usepackage{tikz}
\usepackage{pgfplots}
\usepackage{comment}
\usepackage{acronym}
\usepackage{makecell}

\definecolor{tab101}{RGB}{194,165,207}
\definecolor{tab102}{RGB}{202,0,32}
\definecolor{tab103}{RGB}{0,136,55}
\definecolor{tab104}{RGB}{123,50,148}
\definecolor{tab105}{RGB}{194,165,207}
\definecolor{mygreen}{rgb}{0.69, 0.87, 0.541}%
\definecolor{myblue}{rgb}{0,0.4470,0.7410}
\definecolor{myblack}{rgb}{0.2,0.2,0.2}

\usepackage{bm}
\renewcommand{\vec}[1]{\bm{#1}}
\newcommand{\textsub}[1]{\textnormal{#1}}
\newcommand{\CW}{\mathcal{C}}

\newcommand{\ddesign}{d_{\textnormal{des}}}

\newcommand{\que}{\mathord{?}}
\newcommand{\compl}[1]{\mathord{\sim} #1}

\newcommand{\ZO}{\{0, 1\}}
\newcommand{\ZQO}{\{0, \que, 1\}}

\DeclareMathOperator{\DF}{\mathsf{D}}
\DeclareMathOperator{\BDD}{\mathsf{BDD}}

\newcommand{\DFC}{\DF_{\textnormal{C}}}

\DeclareMathOperator{\dH}{d}

\newcommand{\Sp}{\mathcal{S}}

\newcommand{\EsNO}{\frac{E_{\textnormal{s}}}{N_0}}
\newcommand{\lEsNO}{E_{\textnormal{s}} / N_0}

\DeclareMathOperator{\QFunc}{Q}

\newcommand{\Topt}{T_{\textnormal{opt}}}

\newcommand{\kc}{k_{\textnormal{c}}}

\newcommand{\dnE}[1]{\operatorname{d}_{\compl{\textnormal{E}(#1)}}}

\newcommand{\Es}{E_{\textnormal{s}}}

\newcommand{\No}{N_{\textnormal{0}}}

\usepackage{bm}
\renewcommand{\vec}[1]{\bm{#1}}
\usepackage{bbm}

\newcommand{\ind}[1]{\mathbbm{1}_{\{#1\}}}

\newcommand{\deltaC}{\delta_{\textnormal{c}}}
\newcommand{\epsilonC}{\epsilon_{\textnormal{c}}}

\newcommand{\pone}{\vec{p}_{\textnormal{1}}}
\newcommand{\ptwo}{\vec{p}_{\textnormal{2}}}
\newcommand{\yimpl}{\vec{y}_{j, \textsub{IMP}}^{(\ell)} }
\newcommand{\yemplk}{\vec{y}_{j, \textsub{EMP}}^{(k,\ell)} }

\newcommand{\yimp}{\vec{y}}
\newcommand{\yimpk}{y_{k}}
\newcommand{\yone}{\vec{y}_{\textnormal{1}}}
\newcommand{\ytwo}{\vec{y}_{\textnormal{2}}}
\newcommand{\yonek}{\vec{y}_{\textnormal{1}}^k}
\newcommand{\ytwok}{\vec{y}_{\textnormal{2}}^k}
\newcommand{\yimpi}{\vec{y}_{i}}

\newcommand{\wimp}{\vec{w}_{\textnormal{IMP}}}
\newcommand{\wimpi}{\vec{w}_{\textnormal{IMP},i}}
\newcommand{\wimpk}{w_{\textnormal{IMP}, k}}

\newcommand{\yempk}{\vec{y}^{k} }
\newcommand{\yempki}{\vec{y}_{i}^{k}}
\newcommand{\wemp}{\vec{w}}
\newcommand{\wempk}{w_{k}}

\newcommand{\cone}{\vec{c}_{\textnormal{1}}}
\newcommand{\ctwo}{\vec{c}_{\textnormal{2}}}

\newcommand\blfootnote[1]{%
  \begingroup
  \renewcommand\thefootnote{}\footnote{#1}%
  \addtocounter{footnote}{-1}%
  \endgroup
}

\binoppenalty=\maxdimen
\relpenalty=\maxdimen

\acrodef{HDD}{hard-decision  decoding}
\acrodef{TPD}{turbo product decoding}
\acrodef{iBDD}{iterative bounded distance decoding}
\acrodef{BDD}{bounded distance decoding}
\acrodef{SA-HDD}{soft-aided \ac{HDD}}
\acrodef{PC}{product code}
\acrodef{AD}{anchor decoding}
\acrodef{HRB}{highly-reliable bit}
\acrodef{EaE}{error-and-erasure}
\acrodef{EaED}{\ac{EaE} decoder}
\acrodef{SABM}{soft-aided bit marking}
\acrodef{BI-AWGN}{binary-input additive white Gaussian noise}
\acrodef{BCH}{Bose--Chaudhuri--Hocquenghem}
\acrodef{DRS}{dynamic reliability score}
\acrodef{DRSD}{dynamic reliability score decoder}
\acrodef{BER}{bit error rate}
\acrodef{SABM-SR}{SABM with scaled reliabilities}
\acrodef{NCG}{net coding gain}
\acrodef{EMP}{extrinsic message passing}
\acrodef{IMP}{intrinsic message passing}
\acrodef{LCEA}{low-complexity EMP algorithm}
\acrodef{h-LCEA}{heuristic-LCEA}
\acrodef{SCC}{staircase code}
\acrodef{BSC}{binary symmetric channel}
\acrodef{DE}{density evolution}
\acrodef{LDPC}{low-density parity-check}
\acrodef{GLDPC}{generalized LDPC}
\acrodef{SC-GLDPC}{spatially-coupled \ac{GLDPC}}
\acrodef{VN}{variable node}
\acrodef{CN}{constraint node}
\pgfplotsset{compat=1.16} 
\begin{document}
\def\hw{-0.00001ex}
\title{E\hspace*{\hw}r\hspace*{\hw}r\hspace*{\hw}o\hspace*{\hw}r\hspace*{\hw}-\hspace*{\hw}a\hspace*{\hw}n\hspace*{\hw}d\hspace*{\hw}-\hspace*{\hw}e\hspace*{\hw}r\hspace*{\hw}a\hspace*{\hw}s\hspace*{\hw}u\hspace*{\hw}r\hspace*{\hw}e\hspace*{\hw} D\hspace*{\hw}e\hspace*{\hw}c\hspace*{\hw}o\hspace*{\hw}d\hspace*{\hw}i\hspace*{\hw}n\hspace*{\hw}g\hspace*{\hw} \hspace*{\hw}o\hspace*{\hw}f\hspace*{\hw} \hspace*{\hw}P\hspace*{\hw}r\hspace*{\hw}o\hspace*{\hw}d\hspace*{\hw}u\hspace*{\hw}c\hspace*{\hw}t\hspace*{\hw} a\hspace*{\hw}n\hspace*{\hw}d\hspace*{\hw} S\hspace*{\hw}t\hspace*{\hw}a\hspace*{\hw}i\hspace*{\hw}r\hspace*{\hw}c\hspace*{\hw}a\hspace*{\hw}s\hspace*{\hw}e\hspace*{\hw}\\ Codes with Simplified Extrinsic Message Passing}

 \author{%
   \IEEEauthorblockN{Sisi Miao, Lukas Rapp, and Laurent Schmalen}
   \IEEEauthorblockA{Karlsruhe Institute of Technology (KIT),
Communications Engineering Lab (CEL),
76187 Karlsruhe, Germany\\
Email: {\{\texttt{sisi.miao}, \texttt{schmalen}\}\texttt{@kit.edu}}, \texttt{lukas.rapp3@student.kit.edu}}}

\maketitle
\begin{abstract}
The decoding performance of product codes and staircase codes based on iterative bounded-distance decoding (iBDD) can be improved with the aid of a moderate amount of soft information, maintaining a low decoding complexity. One promising approach is error-and-erasure (EaE) decoding, whose performance can be reliably estimated with density evolution (DE). However, the extrinsic message passing (EMP) decoder required by the DE analysis entails a much higher complexity than the simple intrinsic message passing (IMP) decoder. In this paper, we simplify the EMP decoding algorithm for the EaE channel for two commonly-used EaE decoders by deriving the EMP decoding results from the IMP decoder output and some additional logical operations based on the algebraic structure of the component codes and the EaE decoding rule. Simulation results show that the number of BDD steps is reduced to being comparable with IMP. Furthermore, we propose a heuristic modification of the EMP decoder that reduces the complexity further. In numerical simulations, the decoding performance of the modified decoder yields up to $0.2$\,dB improvement compared to standard EMP decoding.
\end{abstract}

\section{Introduction}
\blfootnote{This work has received funding from the European Research Council (ERC)
under the European Union’s Horizon 2020 research and innovation programme
(grant agreement No. 101001899).}
\Acp{PC}~\cite{Elias1955} and \acp{SCC}~\cite{Smithxx12} are powerful code constructions often used in optical communications. Conventionally, \ac{HDD} of \acp{PC} and \acp{SCC} is based on efficient iterative \ac{BDD} with algebraic component decoders. Recently, to meet the requirements of ultra high-speed optical communications, several hybrid algorithms have been proposed aiming to improve the decoding performance of \ac{HDD} with a certain amount of soft information without increasing the decoding complexity and internal decoder data flow significantly~\cite{sheikh2021refined,sheikh2018low,sheikh2019binary,sheikh2021novel,lei2019improved}.
We focus on a promising approach using \ac{EaE} decoding of the component codes using a 3-level (ternary) channel output. It was shown that EaE decoding improves the coding gain of \ac{PC}s~\cite{Soma2021} and \ac{SCC}s~\cite{Sukmadji2020} based on simulation and stall pattern analysis assuming miscorrection-free decoding. In~\cite{sheikh2021novel}, the performance of EaE decoding is improved with additional miscorrection control. In~\cite{Rapp2021}, we have analyzed the decoding behavior of \ac{EaE} decoding for both \acp{PC} and \acp{SCC} using \ac{DE} formulated for the corresponding \ac{GLDPC} and \ac{SC-GLDPC} ensembles including miscorrections. \ac{DE} has been extensively applied in the analysis of \ac{LDPC} codes and it is well-known that \ac{DE} requires \ac{EMP}. In the context of \ac{PC}s and SCCs, \ac{EMP} also usually yields better decoding performance than simple \ac{IMP}~\cite{Rapp2021}. Moreover, the advantages of EMP compared to IMP for the \ac{BSC} have been shown in~\cite{Jian2017}. The major obstacle in applying \ac{EMP} is that the number of component code decoder executions is proportional to $n^2$ in every half-iteration with $n$ being the block length of the component codes, while \ac{IMP} requires only a linear number of decoding steps. To reduce the complexity, a simplification of \ac{EMP} has been investigated for the \ac{BSC} in~\cite{Jian2017} such that the result of \ac{EMP} decoding can be obtained from the \ac{IMP} decoding result together with some simple logical operations. In this paper, we generalize the simplification of \ac{EMP} in~\cite{Jian2017} to the \ac{EaE} channel to obtain low-complexity \ac{EMP} decoding algorithms with linear complexity. Moreover, we also propose a modification of the algorithm which reduces the complexity even further while delivering improved decoding performance. 

\section{Preliminaries}
\label{sec:prel}
We consider \acp{PC} of rate $r = {\kc^2}/{n^2}$, which can be seen as 2-D arrays of size $n\times n$ where each row/column vector $\vec{x}$ is a codeword of an $(n,\kc,t)$ component code $\mathcal{C}$. $\mathcal{C}$ is either a $(2^{\nu}-1,k_0,t)$ binary \ac{BCH} code or its $(2^{\nu}-1,k_0-1,t)$ even-weight subcode, both able to correct $t$ errors. The designed distance of $\CW$ is denoted as $\ddesign=2t+1$.
An \ac{SCC} of an $(n,\kc,t)$ component code $\mathcal{C}$ and length $L$ consists of a chain of $L$ matrices $\vec{B}_i$ of size $n/2 \times n/2$ where $i\in\{1,2,..,L\}$. Every row of the matrix $[\vec{B}_i^T,\vec{B}_{(i+1)}]$ is a valid codeword of the component code $\CW$.  %
We consider $\CW$ being either shortened \ac{BCH} codes or shortened even-weight \ac{BCH} subcodes. To decode a \ac{PC} or \ac{SCC} codeword, the rows and columns of the blocks are alternately decoded with a component code decoder $\DFC$.

A \ac{PC} can be interpreted as a \ac{GLDPC} code %
and an \ac{SCC} can be viewed as an \ac{SC-GLDPC} code~\cite{Jian2017}. Thus, the decoding performance of \acp{PC} and \acp{SCC} can be predicted via \ac{DE} formulated on an adequate GLDPC or SC-GLDPC ensemble, respectively. A detailed description of constructing the random Tanner graphs of such ensembles can be found in~\cite{Jian2017,Rapp2021, zhang2017spatially}. %

The codewords are transmitted over a \ac{BI-AWGN} channel which outputs $\tilde{y}_i=(-1)^{x_i}+n_i$, where $n_i$ is (real-valued) AWGN with noise variance $\sigma^2 = \frac{1}{2}(\Es/\No)^{-1}$. To obtain the discrete channel output $y_i\in\ZQO$, the values $\tilde{y}_i\in [-T,+T]$ are declared as erasures ``$\que$'', where $T$ is a configurable threshold to be optimized. Values outside this interval are mapped to $0$ and $1$ by the usual \ac{HDD} rule. For a fixed $T$, the capacity of this \ac{EaE} channel is
\begin{equation*}
    \phantom{,}
    C\left(\EsNO, T\right) 
    = 
    c_{\textsub{c}} \log_2\left( \frac{2 c_{\textsub{c}}}{1 - \epsilonC} \right) 
    + \deltaC \log_2 \left( \frac{2 \deltaC}{1 - \epsilonC} \right)
    ,
\end{equation*}
where the probability for an error $\deltaC$ and the probability for an erasure $\epsilonC$ are given by
\begin{equation*}
    \deltaC=\QFunc\left(\sqrt{2\Es/\No} (T + 1)\right)
\end{equation*}
\begin{equation*}
    \epsilonC=1 - \QFunc\left(\sqrt{2\Es/\No} (T - 1)\right) - \QFunc\left(\sqrt{2\Es/\No} (T + 1)\right)
\end{equation*}
and $c_{\textsub{c}} \coloneqq 1 - \deltaC - \epsilonC$.
Numerical optimization of $C(\lEsNO, T)$ with respect to $T$ results in a capacity gain compared to the \ac{BSC} ($T=0$) and an optimal threshold $\Topt$.

\section{Error-and-erasure Decoding}
We consider two commonly used \ac{EaE} decoders as component code decoder $\DFC$. For both algorithms, let $\vec{y} \in \ZQO^n$ be the received row/column vector and define the decoding result $\vec{w}$ as
\begin{equation*}
    \vec{w} \coloneqq  \DFC(\vec{y})\in \CW \cup \{\yimp\},
\end{equation*}
where $\yimp$ is returned unchanged upon decoding failure.

Similar to the Hamming sphere $\Sp_t(\vec{c})$ in $\ZO^n$, we define 
\begin{equation*}
    \Sp^3_t(\vec{c}) \coloneqq \{\vec{y} \in \ZQO^n : 2 \dnE{\vec{y}}(\vec{y}, \vec{c}) + |\operatorname{E}(\vec{y})| < \ddesign(t)\}
\end{equation*}
as the Hamming sphere in $\ZQO^n$, where $|\operatorname{E}(\vec{y})|$ is the number of erasures of $\vec{y}$ and $\dnE{\vec{y}}$ is the Hamming distance at the non-erased coordinates of $\vec{y}$.

The first \ac{EaED} is a modification of~\cite[Sec.~3.8.1]{MoonBook} and is described in Algorithm~\ref{alg:eaed}. 

\begin{algorithm}[t]
	\footnotesize
	\DontPrintSemicolon
\caption{EaED (input: $\vec{y}\in\ZQO^n)$}\label{alg:eaed}
$E\gets$ number of erasures in $\yimp$\;
\lIf(\tcp*[f]{failure}){$E\geq \ddesign$} {$\vec{w}=\yimp$}
\Else{
        $\pone,\ptwo\gets$ two random complementary vectors in $\ZO^{E}$\;
        $\vec{y}_1$,$\vec{y}_2 \in \ZO^n \gets$ $\yimp$ with erasures replaced by $\pone,\ptwo$\;
        \For{$i=1,2$}
            {$\vec{w}_i \gets \BDD(\vec{y}_i)$,$\;\dH_i\gets \infty$\;
            \lIf{$\BDD(\vec{y}_i)\in \CW$}{$\dH_i\gets \dnE{\vec{y}}(\vec{y}, \vec{w}_i)$}
            }
        \lIf(\tcp*[f]{failure}){$\vec{w}_1\not\in \CW,\vec{w}_2\not\in \CW$}{$\vec{w}\gets \vec{y}$}
        \lElseIf{$\vec{w}_i\in \CW,\vec{w}_j\not \in \CW\;(i,j\in\{1,2\},i\neq j)$}{$\vec{w}\gets \vec{w}_i$}
        \Else{
            \lIf{$d_1>d_2$}{$\vec{w}=\vec{w}_2$}
            \lElseIf{$d_2>d_1$}{$\vec{w}=\vec{w}_1$}
            \lElse{$\vec{w}\gets$ random choice from $\{\vec{w}_1,\vec{w}_2\}$}
        }
    }
\end{algorithm}

The second one, referred to as EaED+, is an algebraic \ac{EaE} decoding algorithm that requires only one decoding step~\cite{forney1965decoding}. For EaED+, the decoding result $\vec{w}$ is obtained by
\begin{equation*}
 \vec{w}\coloneqq \DF_{\textsub{EaED+}}(\vec{y})
    =
    \begin{cases}
        \vec{c} & \text{if \(\exists\vec{c}\in\CW\) such that \(\vec{y} \in \Sp_t^3(\vec{c})\)} \\
        \vec{y} & \text{otherwise}
    \end{cases}.
\end{equation*}

Similar to EaED+, $\ac{EaED}$ guarantees that $\vec{w}=\vec{c}$ when there exists a $\vec{c}\in\CW\) such that \(\vec{y} \in \Sp_t^3(\vec{c})$. The difference between both decoders is that EaED, with higher complexity, can sometimes still decode when EaED+ fails (i.e. beyond the designed distance). This subtle difference causes a notable iterative decoding performance gain for \ac{EaED} compared to EaED+~\cite{Rapp2021}.

\section{Message-passing Decoding for \ac{GLDPC} Codes}
\label{sec:message passing}
Iterative decoding of \acp{PC} and \acp{SCC} can be formulated as a message passing decoding process of corresponding \ac{GLDPC} and \ac{SC-GLDPC} codes~\cite{Jian2017}.

We denote by $\nu_{i,j}^{(\ell)}$ the message passed from \ac{VN} $i$ to \ac{CN} $j$ and by $\tilde{\nu}_{i,j}^{(\ell)}$ the message passed from \ac{CN} $j$ to \ac{VN} $i$  in the $\ell$-th iteration. Define $i\coloneqq \sigma_j(k)$ where $k\in \{1,2,\ldots,n\}$ such that $i$ is the index of the \ac{VN} that is connected to the $k$-th socket of the CN $j$. Upon initialization, the outgoing message of a \ac{VN} $i$ is set to the channel output $r_i \in \ZQO$.

In the $\ell$-th \ac{CN} update, each \ac{CN} $j$ receives $n$ incoming messages from all its neighboring \acp{VN}. For \ac{IMP} decoding~\cite{Smithxx12}, the messages are combined into \begin{equation*}
	\yimpl
	\coloneqq
	\left(\nu_{\sigma_j(1), j}^{(\ell)}, \cdots, \nu_{\sigma_j(n), j}^{(\ell)}\right)
\end{equation*}
and are decoded by the component decoder $\DFC\in \{\DF_{\textsub{EaED}},\DF_{\textsub{EaED+}}\}$. The \ac{CN} $j$ then sends the message $\tilde{\nu}_{i,j}^{(\ell)}=[\DFC(\vec{y}_{j, \textsub{IMP}}^{(\ell)})]_k$ back to \ac{VN} $i=\sigma_j(k)$, where $[\vec{y}]_k$ denotes the $k$-th component of vector $\vec{y}$.

For \ac{EMP} decoding~\cite{Jian2017}, one component code decoding is performed to calculate $\tilde{\nu}_{i,j}^{(\ell)}$ for each of the $n$ VNs. For computing $\tilde{\nu}_{i,j}^{(\ell)}$, the $k$-th position of $\yimpl$ is replaced by the channel output $r_i$, yielding
\begin{equation*}
	\yemplk \coloneqq
	\left(
	\nu_{\sigma_j(1), j}^{(\ell)}, \cdots,
	\nu_{\sigma_j(k-1), j}^{(\ell)}, r_{i}, 
	\nu_{\sigma_j(k+1), j}^{(\ell)}, \cdots
	\right).
\end{equation*}
The \ac{CN} $j$ then sends $\tilde{\nu}_{i,j}^{(\ell)}=[\DFC(\yemplk)]_{k}$ to \ac{VN} $i$.

In the \ac{VN} update, each \ac{VN} \(i\) receives two messages from its connected CNs \(j\), \(j^\prime\)
and forwards to each \ac{CN} the message that it has received from the respective other CN:
\(\nu_{i, j^\prime}^{(\ell+1)} = \tilde{\nu}_{i, j}^{(\ell)}\),
\(\nu_{i, j}^{(\ell+1)} = \tilde{\nu}_{i, j^\prime}^{(\ell)}\).

At the end of message passing, each \ac{VN} randomly chooses one of the incoming messages as its final value. If the message is erased, it is replaced by a random binary value.

\Ac{EMP} guarantees the independence of the messages, which enables the \ac{DE} analysis~\cite{Rapp2021}. Moreover, in EMP, the hard channel outputs are used twice: Once as the initial \ac{VN} value and a second time as a replacement of the intermediate \ac{VN} value, avoiding miscorrections to some extent.

\section{Low-complexity EMP Algorithms (LCEAs)}
\label{sec:LCEA}
To fully use the benefits of EMP and of an accurate performance prediction via DE, we need a low-complexity version of the EMP decoder.
As described in Sec.~\ref{sec:message passing}, the number of component code decoding steps of \ac{EMP} is proportional to $n^2$ instead of $n$ as for IMP. This is the major obstacle for \ac{EMP} decoding in practical applications. For the \ac{BSC}, the \ac{EMP} decoding result can be calculated from the \ac{IMP} decoding result and some simple logical operations, such that the complexity becomes comparable to IMP~\cite{Jian2017}. In this section, we show that similar simplifications can be performed for the \ac{EaE} channel.

We consider one \ac{CN} update and try to predict the \ac{EMP} decoding result $\tilde{\nu}_{\sigma_j(k),j}^{(\ell)}$ destined for VN $\sigma_j(k)$. To simplify the notation, we omit the index of \ac{CN} node $j$ and iteration $\ell$ and use the upper-script $k$ to differentiate between IMP- and EMP-associated intermediate vectors and variables during the decoding. For example, $\yimp$ stands for $\yimpl$ and $\yempk$ stands for $\yemplk$. Let $r_k\in\ZQO$ denote the channel output at the $\sigma_j(k)$-th VN.
The \ac{IMP} decoding result is $\wimp\coloneqq \DFC(\vec{y})$ and the final \ac{EMP} decoding result is denoted by $\wemp$. We denote by $w_k$ the $k$-th position of $\wemp$ such that $\tilde{\nu}_{\sigma_j(k),j}^{(\ell)}=w_k$.
Additionally, for \ac{EaED} (Algorithm~\ref{alg:eaed}), let $\yone$ and $\ytwo$ be the vector where the erasure positions in $\yimp$ are replaced by two random complementary vectors $\pone$ and $\ptwo$, and similarly, define $\yonek$ and $\ytwok$ for $\yempk$. Let $y_i^k$ be the $k$-th component of $\vec{y}^k_i$  for $i\in\{1,2\}$. Finally, $\wimpi \coloneqq \BDD(\yimpi)$ and $\vec{w}_i^k \coloneqq \BDD(\yempki)$.

\subsection{Low-complexity \ac{EMP} Algorithm for the \ac{BSC}}
\label{subsec:LCEA_BSC}
We first revisit the low-complexity \ac{EMP} algorithm for the \ac{BSC}~\cite{Jian2017} from a slightly different perspective. The decoding result is $\wimp \coloneqq \BDD(\yimp)\in \CW \cup \{\yimp\}$. We can obtain the distance $\dH(\wimp,\yimp)$ from the \ac{IMP} decoder and we set $\dH(\wimp,\yimp)=\infty$ in case of a decoding failure. For every position $k$, the distance $\dH(\wimp,\yempk)$ can be obtained from $\dH(\wimp,\yimp)$ by
\begin{equation*}
    \dH(\wimp,\yempk)=\dH(\wimp,\yimp)+\begin{cases}
    0&y_k=r_k\\
    -1&y_k\neq r_k,w_{\textnormal{IMP}}=r_k\\
    +1&y_k\neq r_k,w_{\textnormal{IMP}}\neq r_k.\\
    \end{cases}
\end{equation*}

If $\dH(\wimp,\yempk)\leq t$, then $\BDD(\yempk)=\BDD(\yimp)=\wimp$. Thus, $w_k=\wimpk$.

If $\dH(\wimp,\yempk) > t$, there are only two possible results: 1) If $\yempk\in\Sp_t(\vec{c})$ for some $\vec{c}\in \CW$, then it must hold that $c_k=r_k$ and it follows that $w_k=r_k$. 2) If $\yempk\not \in\Sp_t(\vec{c})$ for any $\vec{c}\in \CW$, then $\BDD(\yempk)$ will fail and $w_k=r_k$.

In summary, we have
\begin{equation}
\label{eq:LCEAinBSC}
    w_k=\begin{cases}
    \wimpk&\dH(\wimp,\yempk)\leq t\\
    r_k&\dH(\wimp,\yempk)> t.
    \end{cases}
\end{equation}

\subsection{Low-complexity \ac{EMP} Algorithm with \ac{EaED}}
Now we analyze the relation between $\wimp$ and $\wemp$ for the \ac{EaED} (Algorithm~\ref{alg:eaed}). The goal is to obtain $\vec{w}_i^k$ and the distance $\dnE{\vec{y}^k}(\vec{w}_i^k,\yempk)$ from the \ac{IMP} decoding result and then predict $\wemp$.

We first consider the case when the number of erasures $E$ in $\yimp$ is too large, i.e., $E \geq \ddesign$. As the decoding process will not turn a non-erased bit into an erasure (EaED does not introduce new erasures), the number of erasures $E^k$ in any $\yempk$ is at least $E$. No decoding will happen and $w_{k}=r_k$ for all $k$. Thus, we output $\wemp = \vec{r}$.

For $E<\ddesign$, we observe the following facts for every $k\in\{1,2,\ldots,n\}$, which lead to the low-complexity EMP algorithm:

If $\yimpk=r_k$, we have $\wempk = \wimpk$ since $\yempk=\yimp$.

If $E^k = E + \ind{y_k \neq \que \wedge r_k = \que} \geq \ddesign$, then $\wempk=r_k$.

Since the \ac{EaED} is based on two BDD outcomes, we first obtain the distance $\dH(\wimpi, \yempki) $ from $\dH(\wimpi, \yimpi)$ calculated in the BDD step to predict the result of $\BDD(\yempki)$. We have
\begin{multline}
\label{eqn:distance1}
\dH(\wimpi, \yempki) = \dH(\wimpi, \yimpi) \\
+ \begin{cases}
0 & r_k = \que, y_{i, k} = p_i \\
-1 & r_k = \que, y_{i, k} \neq p_i, w_{\textnormal{IMP}, i, k} = p_i \\
+1 & r_k = \que, y_{i, k} \neq p_i, w_{\textnormal{IMP}, i, k} \neq p_i \\
0 & r_k \neq \que, y_k = r_k \\
-1 & r_k \neq \que, y_k \neq r_k, \wimpk = r_k \\
+1 & r_k \neq \que, y_k \neq r_k, \wimpk \neq r_k,
\end{cases}
\end{multline}
where $p_i\coloneqq [\vec{p}_i]_k$. From \eqref{eq:LCEAinBSC}, we know that $w_{i,k}^k = w_{\textnormal{IMP},i,k}$ if $\dH(\wimpi, \yempki)\leq t$. For $\dH(\wimpi, \yempki)> t$, $w_{i,k}^k=r_k$ if $r_k\neq \que$ and $w_{i,k}^k=p_i$ if $r_k = \que$. We are left to determine which one of the $\vec{w}_i^k$ will be chosen for the following three cases.

\textbf{Case 1}: $\dH(\wimpi,\yempki)\leq t$ for both $i\in\{1,2\}$. In this case, $\vec{w}_i^k =\wimpi$. We just need to compare the distance between $\vec{w}_i^k$ and $\yempk$ at the unerased coordinates of $\yempk$. We observe that
\begin{multline}
\label{eqn:distance2}
\dnE{\yempk}(\yempk, \vec{w}_i^k) 
= \dnE{\yimp}(\yimp, \wimpi) \\ 
+
\begin{cases}
0 & y_k = \que \\
0 & y_k \neq \que, r_k = \que, w_{\textnormal{IMP}, i, k} = y_i^k\\
-1 & y_k \neq \que, r_k = \que, w_{\textnormal{IMP}, i, k} \neq y_i^k \\
0 & y_k \neq \que, r_k \neq \que, y_{i, k} = r_k \\
-1 & y_k \neq \que, r_k \neq \que, y_{i, k} \neq r_k, w_{\textnormal{IMP}, i, k} = r_k\\
+1 & y_k \neq \que, r_k \neq \que, y_{i, k} \neq r_k, w_{\textnormal{IMP}, i, k} \neq r_k,
\end{cases}
\end{multline}
where $\dnE{\yimp}(\yimp, \wimpi)$ for both $i=1$ and $2$ are calculated in the IMP decoding step.
Then we can choose the value of $\wempk$ based on the distance comparison of $\dnE{\yempk}(\yempk, \vec{w}_i^k)$:
\begin{equation}
\label{eqn:wkcase1}
	w_k =
	\begin{cases}
	w_{\textnormal{IMP,1}, k} & \dnE{\vec{y}^k}(\vec{y}^k, \vec{w}_1^k) < \dnE{\vec{y}^k}(\vec{y}^k, \vec{w}_2^k) \\
	w_{\textnormal{IMP,2}, k} &  \dnE{\vec{y}^k}(\vec{y}^k, \vec{w}_1^k) > \dnE{\vec{y}^k}(\vec{y}^k, \vec{w}_2^k).
	\end{cases}
\end{equation}
In the case of equality, one of the $w_{\textnormal{IMP}, i, k}$ is chosen at random.

\textbf{Case 2}: $\dH(\wimpi,\yempki) > t$ for both $i\in\{1,2\}$.

If $r_k\neq \que$, then $w_k=r_k$ as in Sec.~\ref{subsec:LCEA_BSC}.

If $r_k=\que$, $y_k=p_i$ has to hold for one of the $i\in\{1,2\}$ as we have $y_k\neq r_k=\que$. We assume $y_k=p_i$, meaning that both $\vec{y}_i^k=\vec{y}_i$ are not decodeable by BDD. Then we only need to consider $\BDD(\vec{y}^k_j)$, which could either be a failure (then $w_k=r_k$) or a success with the condition $\vec{w}_{j,k}=p_j$ (then $w_k=p_j$). Hence, this case is not deterministic and re-decoding is required, i.e., $w_k = [\DF_{\textsub{EaED}}(\yempk)]_k$ needs to be computed with an actual EaED step. Heuristically, we could set $w_k=r_k$ to avoid the extra decoding step.

\textbf{Case 3}: $\dH(\wimpi,\yempki) \leq t$ and $\dH(\vec{w}_{\textnormal{IMP},j}, \vec{y}^k_{j}) > t$ $(i,j\in\{1,2\},i\neq j)$. This is the most complicated case and $w_k$ is only solvable for some special cases:

\emph{Case 3.1}: If $r_k=w_{\textnormal{IMP},i,k}$, then $w_k=r_k$, because we have $w_{i,k}=r_k$ ($\dH(\wimpi,\yempki) \leq \dH(\wimpi,\yimp) $) and $w_{j,k}=r_k$ (using \eqref{eq:LCEAinBSC}). Both $w_{i,k}$ and $w_{j,k}$ are consistent.

\emph{Case 3.2}: If $(r_k\neq \que,y_{j,k}=r_k)$ or $(r_k=\que,y_{j,k}=p_j)$, then $w_k=w_{\textnormal{IMP},i,k}$ as $\BDD(\vec{y}_j^k)=\BDD(\vec{y}_j)\not\in \CW$ and $\BDD(\vec{y}_i^k)=\BDD(\vec{y}_i)=\wimpi\in\CW$. Note that the first condition implies $y_{j,k}=y_{i,k}=y_k=r_k$, which has already been covered above.

\emph{General case}: $w_k$ is not solvable and an extra decoding is required. As we have $\vec{w}_i=\BDD(\vec{y}_i^k)=\wimpi$, $w_k\neq w_{\textnormal{IMP},i,k}$ if and only if $\vec{w}_j=\BDD(\vec{y}_j^k)\in \CW$ and $\dnE{\yempk}(\yempk,\vec{w}_j)\leq \dnE{\yempk}(\yempk,\vec{w}_i)$. This is possible when $y_{j,k}^k\neq y_{j,k}$ which is true if none of the conditions above holds. A heuristic approach is to set $w_k=w_{\textnormal{IMP},i,k}$ to avoid additional decoding as $\BDD(\vec{y}_j^k)$ is more prone to a miscorrection than $\BDD(\vec{y}_i^k)$.

We call the decoding process described above \ac{LCEA} with EaED. Additionally, \ac{h-LCEA} is a simplified version of \ac{LCEA} where the extra decoding step is avoided by setting $w_k$ with a heuristic value. Both algorithms are summarized in Algorithm~\ref{alg:lcea eaed}. The difference lies on lines 14,15 and lines 19,20.

\SetKw{Continue}{continue}
\begin{algorithm}[t]
	\footnotesize
	\DontPrintSemicolon
\caption{LCEA and h-LCEA with EaED}\label{alg:lcea eaed}
\lIf{$E\geq \ddesign$} {
    $\vec{w}=\vec{r}$, \KwRet{}}
    \lFor{$i=1,2$}{$\wimpi \gets \BDD(\yimpi)$}
    $\wimp \gets \DF_{\textsub{EaED}}(\yimp)$\;
        \For {$k = 1, 2, \dots, n$}{
            \lIf{$y_k=r_k$}{
                $w_k \gets \wimpk$, \Continue
            }
                $E^k \gets E + \ind{y_k \neq \que \wedge r_k = \que} $\;
                \lIf{$E^k\geq \ddesign$}
                    {$\wempk=r_k$, \Continue}
                
                \lFor{$i=1,2$}{$d_i\gets \dH(\wimpi,\yempki)$ using \eqref{eqn:distance1}}
                \uIf(\tcp*[f]{Case 1}){$d_1\leq t, d_2\leq t$}
                {
                calculate $w_k$ using \eqref{eqn:distance2} and \eqref{eqn:wkcase1}
                }
                \uElseIf(\tcp*[f]{Case 3}){$d_i\leq  t, d_j> t$ $(i,j\in\{1,2\},i\neq j)$}{
                \lIf{$((r_k=w_{\textnormal{IMP},i,k})%
                \vee (r_k=\que \wedge y_{j,k}=p_j))$}{$w_k=\wimpk$}
                    \Else{LCEA: $w_k\gets [\DF_{\textsub{EaED}}(\yempk)]_k$\;
                        \textcolor{tab104}{h-LCEA: $w_k\gets \wimpk$}}
                    }
                 \Else(\tcp*[f]{Case 2}){
                     \lIf{$r_k\neq \que$}{$w_k=r_k$}
                     \Else{LCEA: $w_k\gets    [\DF_{\textsub{EaED}}(\yempk)]_k$\;
                     \textcolor{tab104}{h-LCEA: $w_k\gets r_k$}}
                 }
                 
                }

\end{algorithm}

\subsection{Low-complexity \ac{EMP} Algorithm with EaED+}
For EaED+, a similar analysis as for the BSC can be performed with
\begin{equation*}
    \tilde{\dH}(\wimp,\yimp)=2\dnE{\yimp}(\wimp,\yimp)+|\operatorname{E}(\yimp)|,
\end{equation*}
which involves both errors and erasures. We calculate
\begin{multline*}
        \tilde{\dH}(\wimp,\yempk) = \tilde{\dH}(\wimp,\yimp)\\
    +\begin{cases}
    +2&y_k\neq \que, y_k=\wimpk,r_k\neq \wimpk,r_k\neq \que\\
    +1&y_k\neq \que, y_k=\wimpk,r_k= \que\\
    -1&y_k\neq \que, y_k\neq \wimpk,r_k= \que\\
    -2&y_k\neq \que, y_k\neq \wimpk,r_k= \wimpk,r_k\neq \que\\
        0&\textnormal{otherwise.}\\
    \end{cases}
\end{multline*}

If $\tilde{\dH}(\wimp,\yempk)< \ddesign$, then $w_k=\wimpk$. 

If $\tilde{\dH}(\wimp,\yempk)\geq \ddesign$, two cases may occur.

\textbf{Case 1}: If $r_k\neq \que$, then $\DF_{\textsub{EaED+}}(\yempk)$ will either fail ($w_k=r_k$) or succeed with $r_k=w_k$ (Sec.~\ref{subsec:LCEA_BSC}). 

\textbf{Case 2}: $r_k=\que$: If $\DF_{\textsub{EaED+}}(\yempk)$ fails, then $w_k=r_k$. We need to determine if it is possible that $\DF_{\textsub{EaED+}}(\yempk)$ succeeds. 

\emph{Case 2.1}: We first assume that $\DF_{\textsub{EaED+}}(\yimp)$ succeeded. With $\cone=\wimp$, $\tilde{\dH}(\cone,\yimp) = \ddesign-1$. Let $\ctwo \in \CW$ such that $\tilde{\dH}(\ctwo,\yempk)\leq \ddesign-1$. This is only possible when $\tilde{\dH}(\ctwo,\yimp)\leq \ddesign$ and $y_k\neq c_{2,k}$. We can see that $\tilde{\dH}(\cone,\ctwo)\leq \tilde{\dH}(\cone,\yimp)+\tilde{\dH}(\ctwo,\yimp)=2\ddesign-1$ as
\begin{align*} 
    \tilde{\dH}(\cone,\ctwo)  &=2\dH(\cone,\ctwo)\\
    &=2(\dnE{\yimp}(\cone,\ctwo)+\dH_{\operatorname{E}(\yimp)}(\cone,\ctwo))\\
    &\leq 2(\dnE{\yimp}(\cone,\ctwo))+2|\operatorname{E}(\yimp)|\\
    &\leq 2(\dnE{\yimp}(\cone,\yimp)+\dnE{(\yimp)}(\ctwo,\yimp))+2|\operatorname{E}(\yimp)|\\
    &=  \tilde{\dH}(\cone,\yimp)+\tilde{\dH}(\ctwo,\yimp).
\end{align*}
Since $\CW$ is a linear code, we know that $\tilde{\dH}(\cone,\ctwo)\geq 2\ddesign$, resulting in a contradiction. Hence, this is an impossible case. 

\emph{Case 2.2}: If $\DF_{\textsub{EaED+}}(\yimp)$ fails, $\DF_{\textsub{EaED+}}(\yempk)$ can succeed if $\exists \vec{c}\in \CW, \tilde{\dH}(\vec{c},\yimp)=\ddesign$ and $\tilde{\dH}(\vec{c},\yempk)=\ddesign-1$. This is possible when one error position in $\yimp$ is replaced by an erasure. If this happens, then it must hold that $c_k=\bar{y}_k$. Hence, $w_k=\bar{y}_k$.

In summary, for EaED+, we have
\begin{equation*}
    w_k=\begin{cases}
    \wimpk&\tilde{\dH}(\wimp,\yempk)< \ddesign\\
    r_k&(\tilde{\dH}(\wimp,\yempk)\geq \ddesign) \wedge \Bar{\mathsf{a}}\\
    r_k \text{ or } \Bar{y_k}&(\tilde{\dH}(\wimp,\yempk)\geq \ddesign) \wedge \mathsf{a}.
    \end{cases}
\end{equation*}
where $\mathsf{a} \coloneqq \left((r_k=\que )\wedge (y_k \neq r_k )\wedge (\DF_{\textsub{EaED+}}(\yimp)=\text{ fail})\right)$ is a condition. A re-decoding is required for the third case.

\section{Simulation Results}
\label{sec:simulation}

\begin{figure*}
\includegraphics{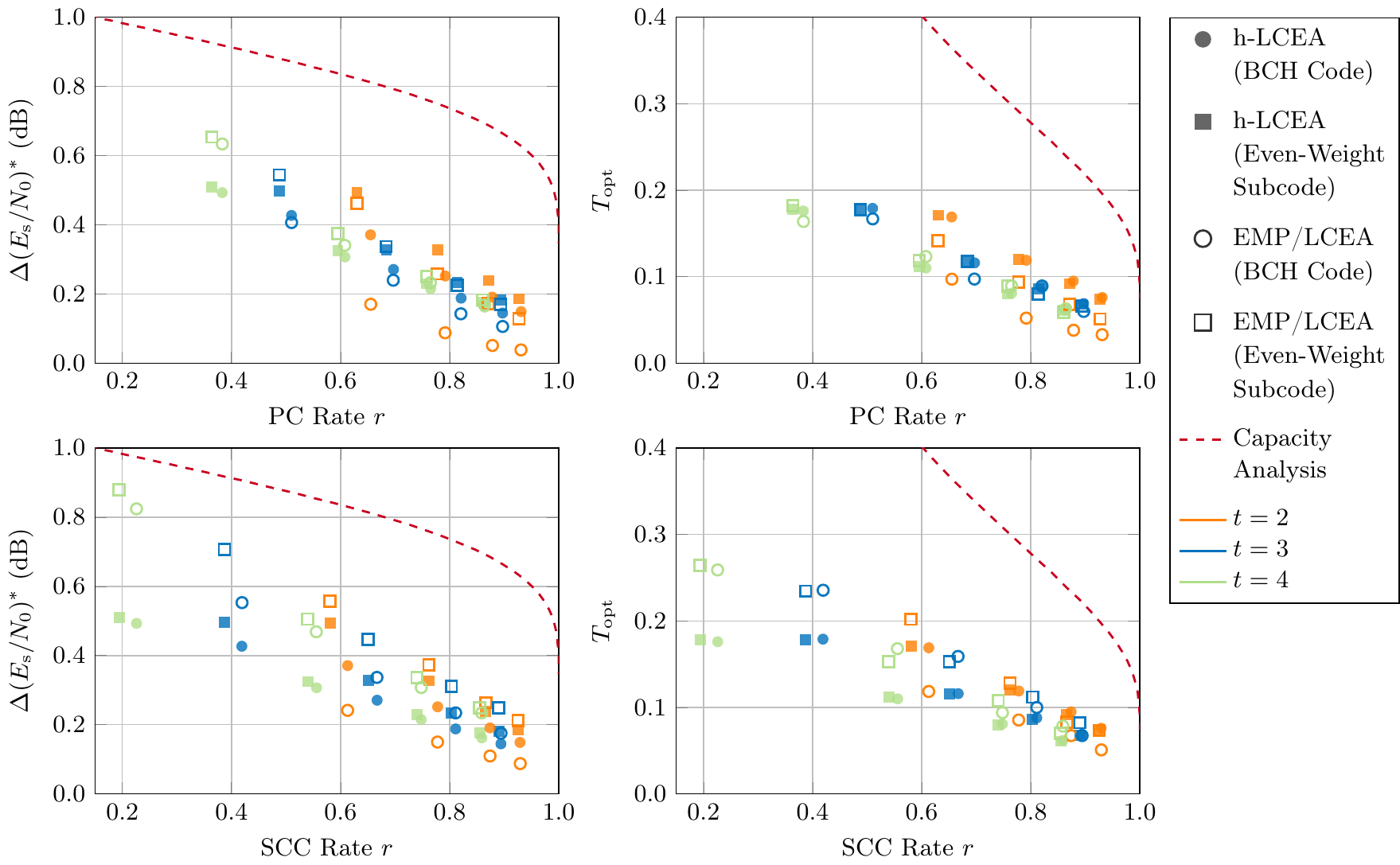}
\vspace{-1ex}
\caption{Noise threshold gain and $\Topt$ for \acp{PC} and \acp{SCC} in the \ac{EaED} decoding with h-LCEA and conventional EMP}
\vspace{-2ex}
\label{plot:gain}
\end{figure*}

\begin{figure}[htbp]
        \centering
    \includegraphics{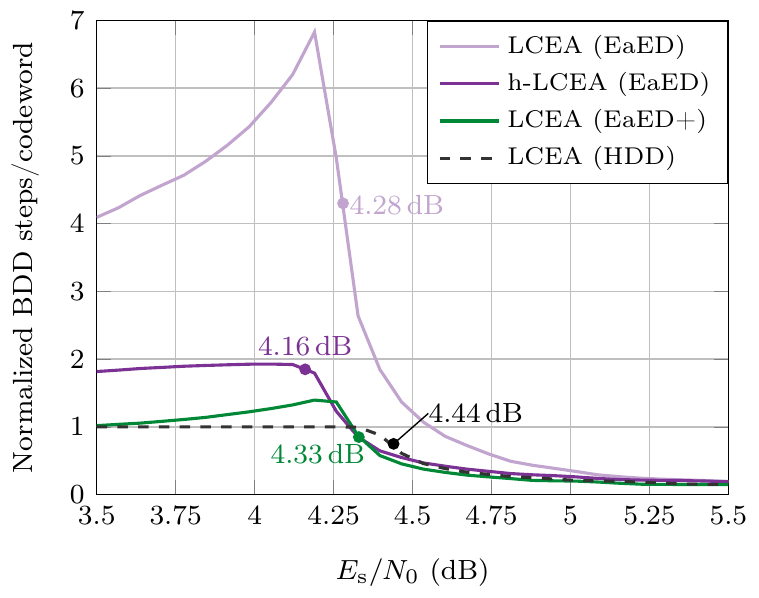}
     	\vspace{-1.5ex}
    	\caption{Normalized number of BDD steps in LCEA-based decoding of $(511,483,3)$ even-weight BCH code-based PC}
     	 \vspace{1ex}
	    \label{plot:complexity}
\end{figure}

We evaluate the performance of h-LCEA with \ac{EaED} by simulation and compare it with conventional \ac{EMP} decoding.
We calculate the noise threshold defined as the minimal $\Es/\No$ with which the target \ac{BER} of $10^{-4}$ after a fixed number of iterations (for \acp{PC} $20$ iterations and for \acp{SCC} $3$ iterations with window length $7$) is achieved numerically by a Monte Carlo approach along with a binary search. The component codes are as described in Sec.~\ref{sec:prel} with parameters $n\in\{63,127,255,511\}$ and $t\in \{2,3,4\}$. We further find the optimal erasure threshold $\Topt$ during the search. The noise threshold difference (gain) $\Delta(\Es/\No)^{*}$ compared to an iterative \ac{HDD} EMP decoder is calculated and shown in Fig.~\ref{plot:gain}.
Additionally, the noise threshold difference for the conventional EMP (implemented using the LCEA) and the respective $\Topt$ are shown. 
The results of LCEA with EaED+ are not shown for the sake of clarity as the EaED+ decoder usually yields smaller performance gains than the \ac{EaED} decoder.
The h-LCEA yields a larger noise threshold gain compared to conventional \ac{EMP} for most of the PCs and some of the SCCs. For small $t$, the gain is relatively large. This is due to the fact that the heuristic value avoids certain miscorrections to some extent, which is particularly beneficial as miscorrections occur approximately with probability $1/t!$ for binary BCH codes~\cite{mceliece1986decoder, justesen2010performance}. The dashed curves mark the maximal achievable gain of the EaE channel compared to a BSC and the corresponding $\Topt$ as described in Sec.~\ref{sec:prel}. Our proposed h-LCEA decoder achieves roughly half of the capacity gain.

As the increased number of component code decoding steps is the major cause of the EMP complexity overhead, we compare the number of BDD steps of the proposed LCEAs based on the EaE decoders and BDD in a $20$ half-iterations decoding for a \ac{PC} constructed with the $(511,3)$ even-weight BCH subcode as an example. The optimal erasure thresholds $\Topt$ are used. 
The results are plotted in Fig.~\ref{plot:complexity} together with their respective noise thresholds for a target BER of $10^{-4}$. By setting $T=0$, the result of LCEA with HDD~\cite{Jian2017}, which is comparable to a normal iterative HDD with IMP, is obtained and used as a baseline. Due to algorithm termination upon decoding success, the curves converge at high $\Es/\No$. The number of BDD steps required for LCEA with \ac{EaED} is still several times higher than for HDD because of the required re-decoding for some bits in \ac{EMP} and the two BDD steps for words with erasures. The complexity can be reduced further with h-LCEA.
For EaED+, the increased complexity for re-decoding is relatively small.

\section{Conclusion}
In this paper, we analyzed \ac{EMP} decoding over the \ac{EaE} channel. This essentially comes down to the question: how will the decoding result change if we change one bit in the vector to be decoded? While this question has a simple and deterministic answer for the BSC with BDD, it is unfortunately not the case for the \ac{EaE} channel due to the uncertainty introduced by the erasures. However, we observe that \ac{EMP} decoding achieves larger coding gains over the \ac{EaE} channel than over the BSC channel. Furthermore, replacing the uncertain result with a value that is more likely to be (closer to) the correct value further improves the decoding performance and reduces the complexity.

\end{document}